\documentclass[preprint,
superscriptaddress,
amsmath,amssymb,aps,showkeys,showpacs,
twoside,final,secnumarabic,
nofootinbib]{revtex4-2}

\usepackage[paperwidth=205mm,paperheight=290mm,top=17mm,bottom=25mm,
inner=17mm,outer=17mm,
twoside]{geometry}

\usepackage{cmap} 
\usepackage[T1,T2A]{fontenc}
\usepackage[utf8]{inputenc}
\usepackage[russian,english]{babel}
\usepackage{color}
\usepackage{graphicx}
\usepackage{dcolumn}
\usepackage{bm} 
\usepackage[unicode=true,colorlinks=true,linkcolor=magenta, urlcolor=blue, citecolor = blue,breaklinks]{hyperref}
\usepackage{multirow}
\usepackage{url}
\usepackage{breakurl}

\newcommand{\be}{\begin{equation}}
\newcommand{\ee}{\end{equation}}
\newcommand{\bea}{\begin{eqnarray}}
\newcommand{\eea}{\end{eqnarray}}

\newcommand{\fb}{\mathfrak{b}}

\newcommand{\fg}{\mathfrak{g}}

\newcommand{\cA}{\cal A}
\newcommand{\cL}{\cal L}

\DeclareGraphicsExtensions{.eps}

\newcount\issue
\newcount\Vol
\newcount\numb
\usepackage{fancyhdr} 
\def\Vol{\textbf{80}}
\def\numb{x}
\begin{document}

\title{ CONFERENCE SECTION \\[20pt]
Jet Quenching in Anisotropic Holographic QCD: \\Probing Phase Transitions and Critical Regions} 

\def\addressa{Steklov Mathematical Institute, Russian Academy of
  Sciences, \\ Gubkina str. 8, 119991, Moscow, Russia  \\}

\author{\firstname{P.S.}~\surname{Slepov}}
\email[E-mail: ]{ slepov@mi-ras.ru }
\affiliation{\addressa}


\begin{abstract}

The jet quenching phenomenon in an anisotropic quark-gluon plasma is studied using gauge–gravity duality. We consider a more general orientation of the contour of a lightlike Wilson loop in the boundary field theory. The Nambu–Goto action for a two-dimensional worldsheet, whose boundary is this contour, is evaluated in a five-dimensional bulk. We present the dependence of the jet quenching parameter on the orientation. Discontinuities of the jet quenching parameter occur at a first-order phase transition, and their magnitude depends on the orientation. These dependencies are observed in holographic models for both light and heavy quarks with nonzero temperature, chemical potential, magnetic field, and spatial anisotropy, supported by an Einstein–dilaton–three-Maxwell action.
\end{abstract}

\pacs{Suggested PACS}\par
\keywords{holography, AdS/QCD, jet quenching, lightlike Wilson loop, phase transition, magnetic catalysis  \\[5pt]}

\maketitle
\thispagestyle{fancy}


\section{Introduction}\label{intro}

Studying the locations of phase transitions in QCD at nonzero temperature and chemical potential is an important task. We propose to consider the jet quenching (JQ) parameter as such a probe. The jets of elementary particles are created in collisions of high-energy particles. Collisions of ultra-relativistic heavy-ion beams create a hot, dense medium comparable to conditions in the early universe. The resulting jets interact strongly with this medium, leading to a significant energy loss known as the JQ. The JQ phenomenon is studied in high-energy heavy-ion collisions, particularly in the context of the quark-gluon plasma (QGP) formation. It refers to the energy loss of high-energy partons (quarks and gluons) as they traverse the QGP, a hot and dense medium created in such collisions. The JQ parameter $\hat{q}$ quantifies the average transverse momentum squared transferred from the parton to the medium per unit path length. It is a key observable for understanding the properties of the QGP. This parameter characterizes how energetic partons lose energy via medium-induced gluon radiation. Crucially, $\hat{q}$ links microscopic parton-medium interactions to observable jet suppression. For a quark with energy $E$, the average energy loss scales as $\langle \Delta E \rangle \propto \alpha_s  \hat{q}  L^2 $,
where $L$ is the medium length. 
\\

A non-perturbative calculation scheme for $\hat{q}$ using the holographic approach was proposed in  \cite{1}. They related $\hat{q}$ to a lightlike Wilson loop in the adjoint representation (denoted $W^A$) of $\mathcal{N}=4$ SYM theory:  \bea \langle W^A(\mathcal{C}) \rangle \sim e^{-\frac{1}{4\sqrt{2}} \hat{q}  L^{-} L_{\perp}^2}, \eea
where $\mathcal{C}$ is a rectangular loop with light-cone ($L^{-}$) and transverse ($L_{\perp}$) extensions. Subsequent holographic calculations using the Nambu-Goto action in AdS-Schwarzschild spacetime established $\hat{q}$'s explicit dependence on the 't Hooft coupling $\lambda$ and horizon position $z_h$  \cite{2, 3, 4}.\\

In this paper, we study the JQ parameter $\hat{q}$ dependence on the orientation of the Wilson loop contour in a background with different types of anisotropy. Isotropic and anisotropic models for heavy and light quarks were studied previously in the literature \cite{13, 14, 12, 5, 11,8,  15, 10,16, 6,9, 7, 12} (see also references therein). These models reproduce the position of the phase transition at zero chemical potential expected from lattice calculations (the Columbia plot) \cite{17}. We consider holographic models for heavy and light quarks with non-zero temperature, chemical potential, external magnetic field, and anisotropy parameter $\nu$, which reproduce the magnetic catalysis effect \cite{13, 14}.   Special attention is paid to $\hat{q}$'s behavior near phase transitions: it varies smoothly across second-order transitions but exhibits discontinuities at first-order transitions. We compare our results with those in \cite{19, 20,18}.
\\

The paper is organized as follows. In Sect. \ref{2} we consider the Nambu–Goto action for the string configuration along the horizon and present an analytical expression for the JQ at an arbitrary angle. In Sect. \ref{3} the JQ is calculated for light and heavy quarks models. Subsect. \ref{3.1} describes the holographic models for light and heavy quarks. In Subsect. \ref{3.2} numerical results for the JQ parameter for different orientations are presented. Finally, in Sect. \ref{4} the main results are summarized.

\section{Jet quenching for arbitrary orientation  in anisotropic background: analytical formula}\label{2}

Let us consider a fully anisotropic background in the string frame with a warp factor $\fb_{s}(z)=\fb(z) e^{\sqrt{2/3}\,\phi(z)}$, where $\phi(z)$ is a dilaton field and $\fb(z)$ is the warp factor in the Einstein frame:
\bea
  &&ds^2 = G_{\mu\nu}dx^{\mu}dx^{\nu}= \label{Gbackgr}   \cfrac{L^2 \fb_{s}(z)}{z^2} \left[
    - \, g(z) dt^2 + \fg_1 (z) dx_1^2 + \fg_2 (z) dx_2^2 + \fg_3 (z) dx_3^2 +
    \cfrac{dz^2}{g(z)} \right]. 
\eea 
Here $L$ is AdS radius, $g(z)$ is a blackening function, $\fg_1 (z)$, $\fg_2 (z)$, $\fg_3 (z)$ are anisotropic functions; $z$ is a holographic coordinate, and we consider the interval $ 0 \le  z \le  z_h$, where $z=0$ is the boundary and $z=z_h$ is the horizon. The functions $\fb_s(z)>0$, $\fg_1 (z)>0$, $\fg_2 (z)>0$, $\fg_3 (z)>0$, and $g(z) \ge 0$  on  the interval $ 0 <  z \le  z_h$.

Following the holographic approach, we compute the Nambu–Goto action for a probe string in the background  \eqref{Gbackgr}:
\be
  S = \frac{1}{2 \pi \alpha'} \int d\tau d\xi \sqrt{- \det
    h_{\alpha\beta}},
\ee
where the induced metric is $h_{\alpha\beta} = G_{\mu\nu} \partial_{\alpha}
  x^{\mu} \partial_{\beta} x^{\nu}$,
which depends on the string orientation.

The Nambu-Goto action for lightlike, spatial and  temporal Wilson loops,  as well as the holographic entanglement entropy  \cite{21,21,22,23,24,25}, are particular
cases of a Born–Infeld type action:
\be
\label{BI}
{\cal S}=\int _{-\ell/2}^{\ell/2} M(z(\xi))\sqrt{{\cal {\cal
      F}}(z(\xi))+(z^{\prime}(\xi))^ 2} \, d\xi.
\ee
This action defines a 1-dimensional dynamical system with the dynamic variable $z =
z(\xi)$ and "time"{}~$\xi$. Note that the asymptotic behavior of  ${\cal S}$  for large and small character lengths $\ell$ depends sensitively on the properties of the functions $M(z)$ and ${\cal F}(z)$,  as well as on their combination ${\cal V}(z(\xi))\equiv M(z(\xi))\sqrt{{\cal F}(z(\xi))}$ which we call the effective potential. The functions $M(z)$ and ${\cal F}(z)$ are determined by the functions
$\fb_s(z)$, $\fg_1 (z)$, $\fg_2 (z)$, $\fg_3 (z)$, and  $g(z)$ from the metric ansatz \eqref{Gbackgr}, together with the orientation of the worldsheet (surface) in the bulk.  We assume  $M(z)\ge 0$ and ${\cal F}(z) \ge 0$ for  $ 0 <  z < z_h$.
Examples of the functions $M(z)$, ${\cal F}(z)$ were given in \cite{21,21,22,23,24,25}, the specific forms for lightlike Wilson loops are presented in \eqref{MF} and \eqref{MFtheta}.

This system has the first integral:
\bea
\label{FI}
\frac{M(z(\xi)){\cal F}(z(\xi))}{\sqrt{{\cal
      F}(z(\xi))+(z'(\xi))^2}}= P. 
\eea
From \eqref{FI} we can express ${z'(\xi)}$:

\bea
\label{zprime}
z'(\xi)=
\frac{M(z(\xi)) {\cal F}(z(\xi))}{P}\sqrt{1-P^2/(M^2(z(\xi)){\cal F}(z(\xi)))}. 
\eea

\begin{figure}[t!]
  \centering 
  \includegraphics[scale=0.27]{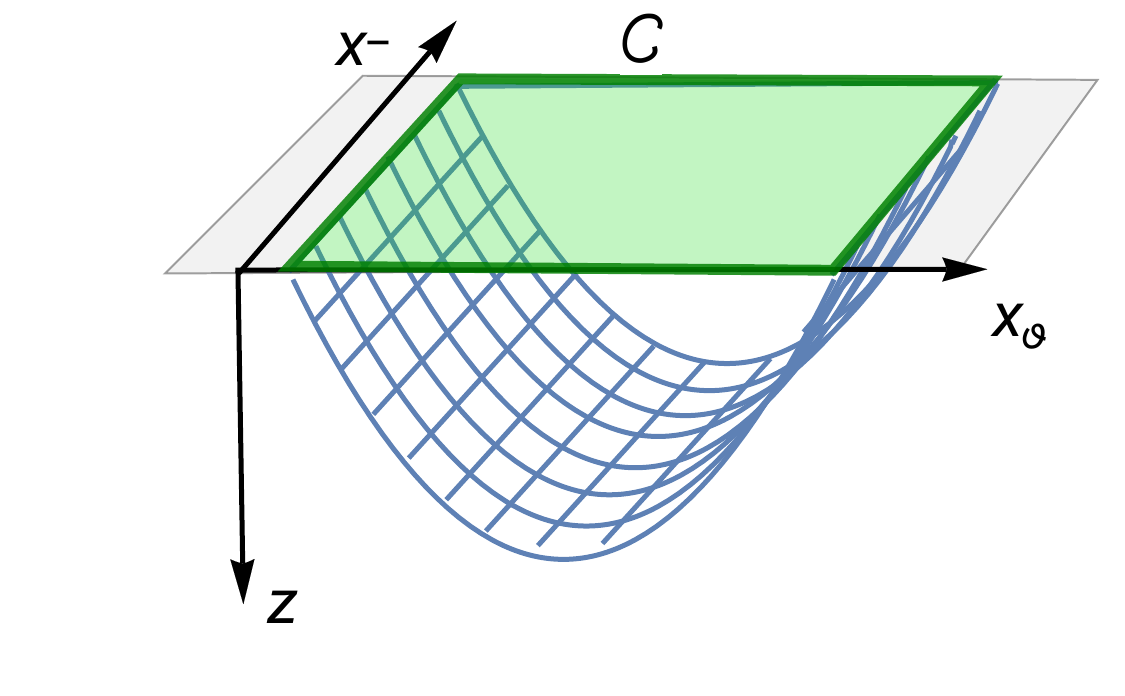}\qquad
  \includegraphics[scale=0.15]{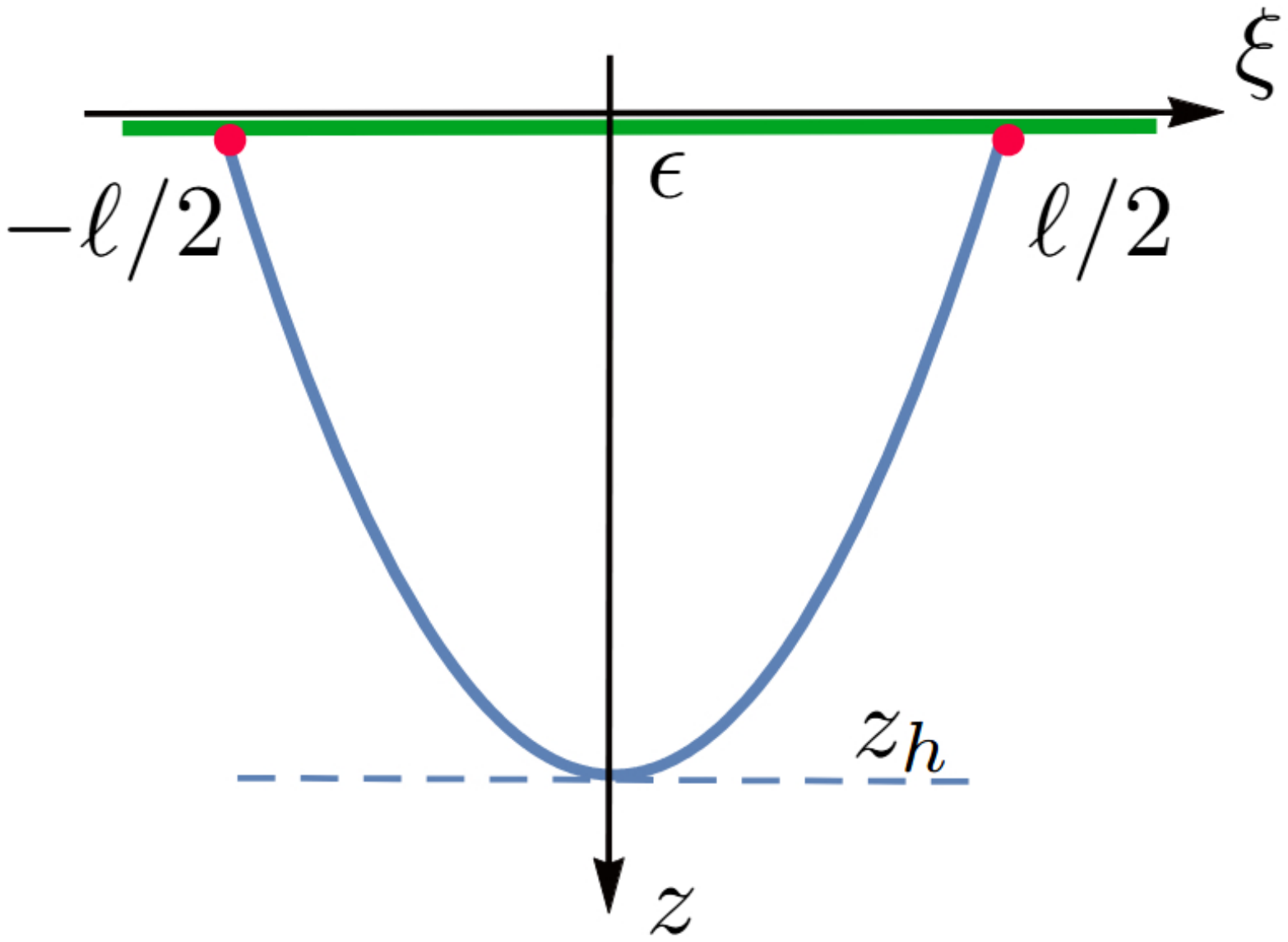} \\ 
A \qquad \qquad  \qquad  \qquad \qquad  \qquad B
  \caption{ \label{WL} A) The Wilson loop contour and the worldsheet. B) The profile of the string configuration along the horizon.}
\end{figure}

Let us consider the string configuration along the horizon $z_h$ (Fig. \ref{WL}). Finding  $z^{\prime}$ from \eqref{FI} one gets  for
the length $\ell$ and the action ${\cal S}$ \eqref{BI}:
\bea\label{ell1}
\frac\ell2 &=&
\int_0^{z_h} \frac{dz}{z'}= \int_0^{z_h} \frac {Pdz} {M(z) {\cal F}(z) \sqrt{1-P^2/(M^2(z){\cal F}(z))}},
\eea
\bea
\frac{{\cal S}}{2} &=& \int_0^{z_h}\frac{M(z)dz}{\sqrt{1-P^2/(M^2(z){\cal F}(z))}}.
\eea

Let us consider small $P$, which corresponds to the small characteristic length $\ell$. Assume $P^2/(M^2(z) {\cal F}(z)) << 1$. To the first order in $P$, $\ell$ becomes 
\bea
\frac{\ell_{(1)}}{2} &=& P\int_0^{z_h} \frac {dz} {M(z) {\cal F}(z)}. 
\eea
For ${\cal S}$ we obtain the following expression up to the second order by $P$
\bea
\frac{{\cal S}}{2} = \int_0^{z_h}M(z)dz + \frac{P^2}{2}\int_0^{z_h}  \frac{dz}{ M(z){\cal F}(z)}+\mathcal{O}(P^3).
\eea
In this case, we can write the following relation between the second order of ${\cal S}$ and the first order of $\ell$:
\bea
{\cal S}_{(2)} = P^2\int_0^{z_h}  \frac{ dz}{ {\cal F}(z) M(z)}=\frac{1}{8\sqrt{2}}\hat{q} L^{-} \ell_{(1)}^2.
\eea

Let us introduce the parameter $a$, which is the inverse of $\hat q$:
\bea 
\hat q=\frac{L^2}{\pi \alpha' a}, \qquad 
a &=&\frac{L^2 L^{-}}{2\sqrt{2} \pi \alpha'}\int_0^{z_h}  \frac{ dz}{ {\cal F}(z) M(z)}.
\eea

For the lightlike Wilson loop in the background \eqref{Gbackgr}, we use the parametrization

\begin{equation}
\label{eqb5}
\ dt=\frac{dx^{+} + dx^{-}}{\sqrt{2}}, \ dx_{i}=\frac{dx^{+} - dx^{-}}{\sqrt{2}}.
\end{equation}

Let us consider the first example of the contour ${\cal C}$ orientation. The rectangular contour ${\cal C}$ consists of
 ``short sides'' of the length $\ell$ along the $x_j$ (that is not $x_i$) direction and ``long
sides''  of the length $L^{-}$ along the $x^{-}$ direction 
\be
x^{-}= \tau,\qquad 
x_j= \xi, \qquad 
x^{+}= const, \qquad 
z=z(\xi).
\ee

The Nambu -Goto action is as follows:

\be
{\cal S}_{j} = \frac{L^2 L^{-}}{2 \pi \alpha'}  \int_{-{\ell \over 2}}^{{\ell \over 2}} d{\xi} \,
\frac{\fb_s(z)}{z^2} \sqrt{\frac{\fg_i(z)-g(z)}{2} \left(\fg_j(z)+\frac{z'^2}{g(z)}\right)}, \quad j\neq i
\ee
where

\bea
M(z)=\frac{L^2 L^{-}}{2 \pi \alpha'}\frac{\fb_s(z)}{z^2} \sqrt{\frac{\fg_i(z)-g(z)}{2 g(z)}}, ~~~~~~{\cal F}(z)=g(z)\fg_j(z). \label{MF} \eea

The expression for the JQ parameter is

\bea
a_{j} &=&\int_0^{z_h}  \frac{ z^2 dz}{\fg_j(z) \fb_s(z) \sqrt{(\fg_i(z)-g(z))g(z)}}, \label{aj}
\eea
which depends on the choice of the direction $x_j$.

Let us consider the second choice of the contour ${\cal C}$ orientation. The rectangular contour ${\cal C}$ consist of 
 ``short sides''  with the length $\ell$ along the $\xi$ and the ``long sides''  with the length $L^{-}$ along the $x^{-}$ direction (Fig. \ref{WL}). To describe the embedding of the 2-dimensional worldsheet in the 5-dimensional spacetime, we use
\be
x^{-}= \tau,\qquad 
 x_j= \xi \sin \theta,  \qquad 
 x_k= \xi \cos \theta, \qquad x^{+}= const, \qquad 
z=z(\xi),
\ee
\be
i,j,k \quad \text{are distinct and} \quad i,j,k=1,2,3.
\ee

The Nambu-Goto action is
\bea
 {\cal S}_{\theta} = \frac{L^2 L^{-}}{ 2\pi \alpha'}  \int_{-{\ell \over 2}}^{{\ell \over 2}} d{\xi} \,
\frac{\fb_s(z)}{z^2} \sqrt{\frac{\fg_i(z)-g(z)}{2} \left(\fg_j(z) sin^2\theta+\fg_k(z) cos^2\theta + \frac{z'^2}{g(z)}\right)},
\eea where
\bea
M(z)=\frac{L^2 L^{-}}{ 2\pi \alpha'} \frac{\fb_s(z)}{z^2} \sqrt{\frac{\fg_i(z)-g(z)}{2 g(z)}}, \qquad \qquad
{\cal F}(z)=g(z)(\fg_j(z) sin^2\theta+\fg_k(z) cos^2\theta). \label{MFtheta}
\eea
The JQ parameter is
\bea
a_{\theta} &=&\int_0^{z_h}  \frac{ z^2 dz}{(\fg_j(z) sin^2\theta+\fg_k(z) cos^2\theta) \fb_s(z) \sqrt{(\fg_i(z)-g(z))g(z)}}. \label{atheta} 
\eea

This result is more general than \eqref{aj} and particular cases of \eqref{atheta} reproduce the previously obtained results for the JQ in anisotropic and isotropic holographic models \cite{20,26,18}. Note that $a_{\theta}$ in \eqref{atheta} is a function on the horizon $z_h$. If the background exhibits a first-order phase transition between small and large black holes (BB) with horizons  $z_{h,1}$ and $z_{h,2}$  respectively, then  $a_{\theta}$  shows a jump between $a_{\theta}(z_{h,1})$ and $a_{\theta}(z_{h,2})$. The temperature at the first-order phase transition satisfies $T_{BB}=T(z_{h,1})=T(z_{h,2})$.

\section{Jet quenching in anisotropic holographic model for heavy and light quarks: numerical results} \label{3}
\subsection{The anisotropic holographic model for heavy and light quarks} \label{3.1}
We consider the following  Lagrangian in the Einstein frame: 
\begin{gather}
  {\cL} = \sqrt{-g} \left[ R 
    - \cfrac{f_0(\phi)}{4} \, F_0^2 
    - \cfrac{f_1(\phi)}{4} \, F_1^2
    - \cfrac{f_3(\phi)}{4} \, F_3^2
    - \cfrac{1}{2} \, \partial_{\mu} \phi \, \partial^{\mu} \phi
    - V(\phi) \right], \label{eq:2.01}  \\
  \phi = \phi(z), \label{eq:2.02} \\
  \begin{split}
    \mbox{ electric  ansatz:} \quad
    &A_0 = A_t(z), \quad A_{i = 1,2,3,4} = 0, \\
    \mbox{magnetic ansatz:} \quad
    &F_1 = q_1 \, dx^2 \wedge dx^3, \quad 
    F_3 = q_3 \, dx^1 \wedge dx^2\, , 
  \end{split}\label{eq:2.03}
\end{gather}
where $\phi$ is the scalar field, $f_0(\phi)$, $f_1(\phi)$
and $f_3(\phi)$ are the gauge kinetic functions associated with Maxwell
fields $F_0$, $F_1$ and $F_3$ respectively, $q_1$ and $q_3$ are
 ``charges'' and $V(\phi)$ is the scalar field potential.  We considered $F_0$, $F_1$ and $F_3$ as first, second and third Maxwell fields, respectively. Note that one may also include an additional Maxwell field $F_2$ with magnetic ansatz \cite{27}. In that case, the holographic model contains three distinct magnetic ansatz. The first Maxwell field sets up a finite non-zero chemical potential, the second Maxwell field provides the spatial anisotropy to reproduce the total multiplicity dependence on energy, and the third Maxwell field supports the magnetic field.

 The metric ansatz is
\begin{gather}
  ds^2 = \cfrac{L^2}{z^2} \ \fb(z) \left[
    - \, g(z) \, dt^2 + dx_1^2 
    + \left( \cfrac{z}{L} \right)^{2-\frac{2}{\nu}} dx_2^2
    + e^{c_B z^2} \left( \cfrac{z}{L} \right)^{2-\frac{2}{\nu}} dx_3^2
    + \cfrac{dz^2}{g(z)} \right] \! , \label{eq:2.04} 
\end{gather}
where $L$ is the AdS-radius, $\fb(z)=e^{2 {\cA}(z)}$ is the warp factor expressed through scale factor ${\cA}(z)$, $g(z)$ is the blackening function, $\nu$ is the primary anisotropy parameter, caused by the non-symmetry of heavy-ion collisions, and $c_B$ is the coefficient of secondary anisotropy related to
the magnetic field $F_3$. The anisotropic functions are $\fg_1 (z)=1$, $\fg_2 (z)=(z/L)^{2-2/\nu}$, $\fg_3 (z)=e^{c_B z^2}(z/L)^{2-2/\nu}$. We set the gravitational constant $8 \pi G_5=1$.

Varying the Lagrangian (\ref{eq:2.01})  with respect to the metric, we obtain the Einstein equations:
\begin{gather}
 R_{\mu \nu} - \cfrac{1}{2} \, g_{\mu \nu} R=T_{\mu \nu}, \quad
  \cfrac{\delta S_m}{\delta g^{\mu \nu}} = \frac{1}{2} \, T_{\mu \nu}
  \sqrt{-g}, \label{eq:2.06}
\end{gather}
and varying with respect to the fields gives the field equations
\begin{gather}
 -\nabla_{\mu}\nabla^{\mu} \phi  + V'(\phi) + \!\! \sum_{i=0,1,3} \!\!
  \cfrac{f_i'(\phi)}{4} \, F_{(i)}^2 = 0, \label{phiEOM} \\
  \partial_{\mu} \left( \sqrt{-g} \, f_i \, F_{(i)}^{\mu \nu} \right)
  = 0, ~~ i=0,1,3. \label{EMEOM}
\end{gather}

The choice of ${\cA}(z)$ distinguishes the heavy and light quarks models: ${\cA}(z) = - \, c z^2/4 \, -  (p-c_B \, q_3)
    z^4$ was considered for heavy
quarks to reproduce magnetic catalysis \cite {14} and previously for particular cases of this modified factor in \cite {14,12,5,7,9,21} and ${\cA}(z) = - \, a\, \ln (b z^2 + 1)$ for light quarks \cite{11,13,8}. With these warp factors, the effect of magnetic catalysis/inverse magnetic catalysis for heavy and light quarks models was reproduce correspondingly. The following parameter values are used for the light quarks model:  $a = 4.046$, $b = 0.01613$, $c = 0.227$; for heavy quarks model  $c = 4 R_{gg}/3$, $R_{gg} = 1.16$, $p = 0.273$, $q_3=5$. After substituting the ansatz into the equations, we solve them and determine the unknown functions $g$, $A_{t}$, $\phi$, $f_1$, $f_3$, and $V$ using the potential reconstruction method \cite{12,14,13}. Note that during the solution we do not obtain  $f_i(\phi)$, $i=0,1,3$ and $V(\phi)$ as explicit functions of $\phi$, but obtain $f_i(\phi(z))$ and $V(\phi(z)) = V(z)$. We then reconstruct $f_i(\phi)$ and $V(\phi)$ using $z =  z(\phi)$ dependence from $\phi(z)$-expression. 

To solve the EOMs we impose the standard boundary conditions on the time component of the first gauge field $A_t$ and on the blackening function $g$:
\bea
A_t(0)=\mu, \quad  A_t(z_h)=0,\quad g(0)=1, \quad g(z_h)=0.
\eea
The physical boundary conditions for the dilaton field in the light and heavy quarks models are

\be
z_{0, LQ}=10 e^{-z_h/4}+0.1 \, ,
\label{LQ-nbc} \quad
z_{0, HQ}=e^{-z_h/4}+0.1 \,,
\ee
where $\phi(z_0)=0$. This choice of boundary conditions reproduces lattice results for the temperature dependence of the string tension (see \cite{28,11,16} and references therein).

The dilaton field for light and heavy quarks is
    \bea
\phi(z)=\int_{z_0}^z\sqrt{-\frac{3 \fb''}{\fb} + \cfrac{9 (\fb')^2}{2 \fb^2} - \cfrac{6 \fb'}{\fb z}
  + \cfrac{4 }{ \nu z^2} \left(
    1 - \cfrac{1}{\nu}
    + \left( 1 - \cfrac{3 \nu}{2} \right) c_B z^2
    - \cfrac{\nu c_B^2 z^4}{2}
  \right)}.  
    \eea

We obtain the following expressions for the blackening functions for the heavy quarks  (HQ) and for the light quarks (LQ) models:
\begin{gather}
  g_{HQ}(z) = e^{c_B z^2} \left[ 1 - \cfrac{\Tilde{I}_1(z)}{\Tilde{I}_1(z_h)}
    + \cfrac{\mu^2 \bigl(2 R_{gg} + c_B (q_3 - 1) \bigr)
      \Tilde{I}_2(z)}{L^2 \left(1 - e^{(2 R_{gg}+c_B(q_3-1))\frac{z_h^2}{2}}
      \right)^2} \left( 1 - \cfrac{\Tilde{I}_1(z)}{\Tilde{I}_1(z_h)} \,
      \cfrac{\Tilde{I}_2(z_h)}{\Tilde{I}_2(z)} \right)
  \right], \label{eq:4.42} \\
  \Tilde{I}_1(z) = \int_0^z
  e^{\left(2R_{gg}-3c_B\right)\frac{\chi^2}{2}+3 (p-c_B \, q_3) \chi^4}
  \chi^{1+\frac{2}{\nu}} \, d \chi, \\  
  \Tilde{I}_2(z) = \int_0^z
  e^{\bigl(2R_{gg}+c_B\left(\frac{q_3}{2}-2\right)\bigr)\chi^2+3 (p-c_B
    \, q_3) \chi^4} \chi^{1+\frac{2}{\nu}} \, d \chi. \label{eq:4.43} 
\end{gather}

\begin{gather}
  g_{LQ}(z) = e^{c_B z^2} \left[ 1 - \cfrac{I_1(z)}{I_1(z_h)}
    + \cfrac{\mu^2 \, (2 c - c_B) \, I_2(z)}{L^2 \left( 1 -
        e^{\left(2c-c_B\right)z_h^2/2} \right)^2} \left( 1 -
      \cfrac{I_1(z)}{I_1(z_h)} \, \cfrac{I_2(z_h)}{I_2(z)} \right)
  \right], \label{eq:2.20}
\end{gather}
where 
\begin{gather}
 I_1(z) = \int_0^{z} \left(1 + b \chi^2 \right)^{3a} e^{- 3 c_B
      \chi^2/2} \, \chi^{1+\frac{2}{\nu}} \, d \chi,
  ~~I_2(z) = \int_0^{z} \left(1 + b \chi^2 \right)^{3a}
  e^{\left(c-2c_B\right)\chi^2} \, \chi^{1+\frac{2}{\nu}} \, d
  \chi. \label{eq:2.22}
\end{gather} 
The blackening functions have similar forms but yield to qualitatively different phase-structure behavior.
For the metric (\ref{eq:2.04}) the temperature, entropy and free energy densities can be written as:
\begin{gather}
 T = \cfrac{|g'|}{4 \pi},~~  s = \cfrac{\sqrt{g_{x_1x_1} \,g_{x_2x_2} \, g_{x_3x_3} }}{4 G_5} \,
  \Bigl|_{z=z_h},~~    F = - \int s \, d T = \int_{z_h}^{\infty} s \, T' dz.
  \label{eq:4.37}
\end{gather}

For these light and heavy quarks models we obtain the following expression for the JQ parameter: 

\bea
a_{\theta} &=&\int_0^{z_h}  \frac{ e^{-2 {\cal A}_s(z)} z^2 dz}{\left((z/L)^{2-2/\nu} sin^2\theta+(z/L)^{2-2/\nu} e^{ \, c_Bz^2} cos^2\theta\right) \sqrt{(1-g(z))g(z)}}.
\eea
Here the scale factor in the string frame ${\cal A}_s(z) ={\cal A}(z)+\sqrt{\frac{1}{6}}\phi(z).$ Note that, for the   cases considered, the effective potential ${\cal V}(z)$ is a monotonically increasing function and the configuration with a dynamical wall is unstable.
The two particular cases of  angles $\theta=0$ and $\pi/2$  are considered in \cite{26}.  

\begin{figure}[h!]
  \centering 
  \includegraphics[scale=0.27]{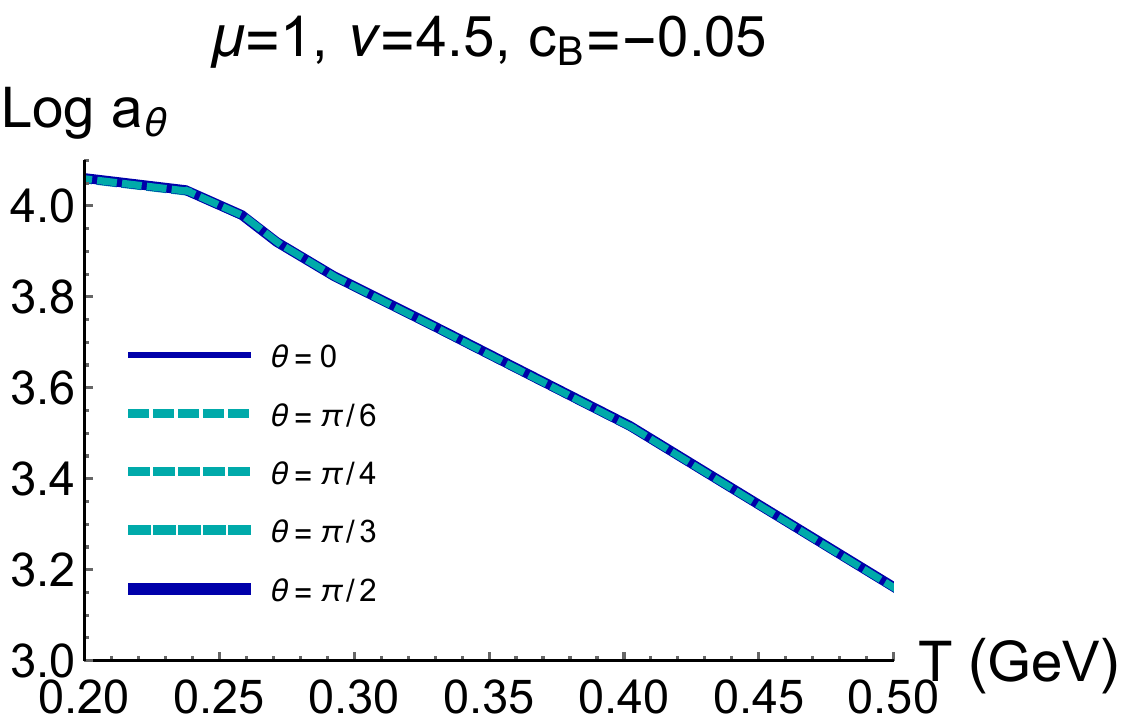}\qquad
  \includegraphics[scale=0.29]{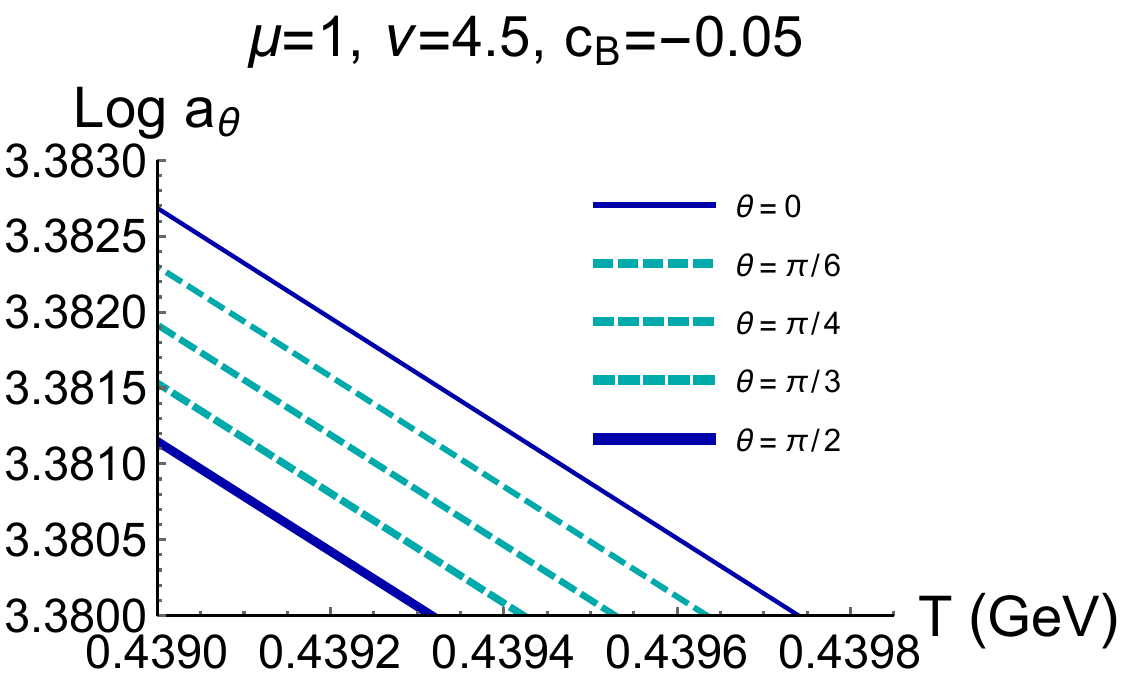} \\ 
A \qquad \qquad  \qquad \qquad \qquad \qquad \qquad \qquad B
  \caption{A) The dependence of the JQ parameter on the temperature for different orientation $\theta = 0, \pi/6, \pi/4, \pi/3, \pi/2$ for $\mu=1$, $\nu=4.5$ and $c_B=-0.05$ in heavy quarks model. B) The zoom view of (A). \label{Htheta1}}
\end{figure}

\begin{figure}[h!]
  \centering 
  \includegraphics[scale=0.24]{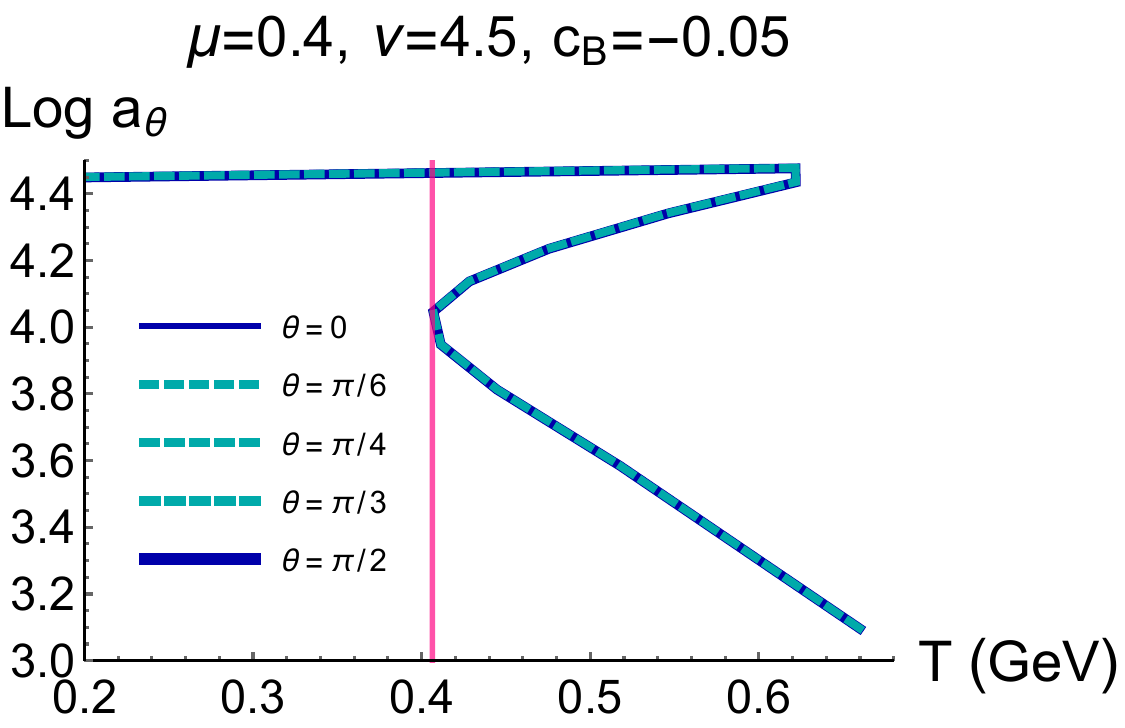}\qquad
  \includegraphics[scale=0.24]{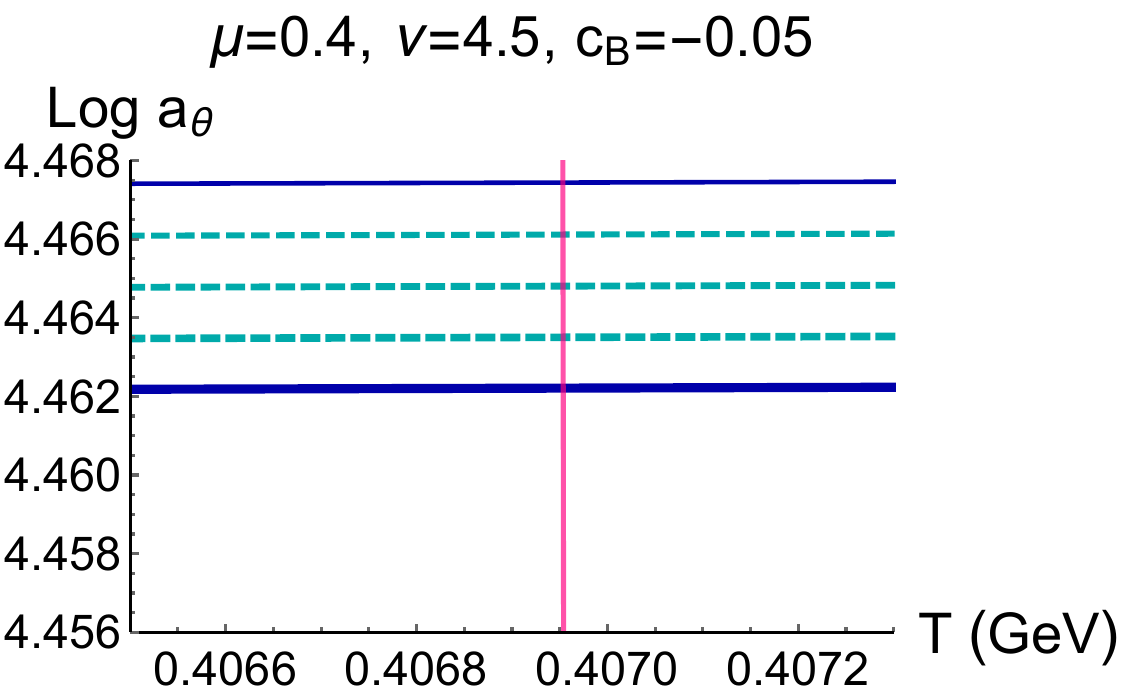}  \quad  \includegraphics[scale=0.24]{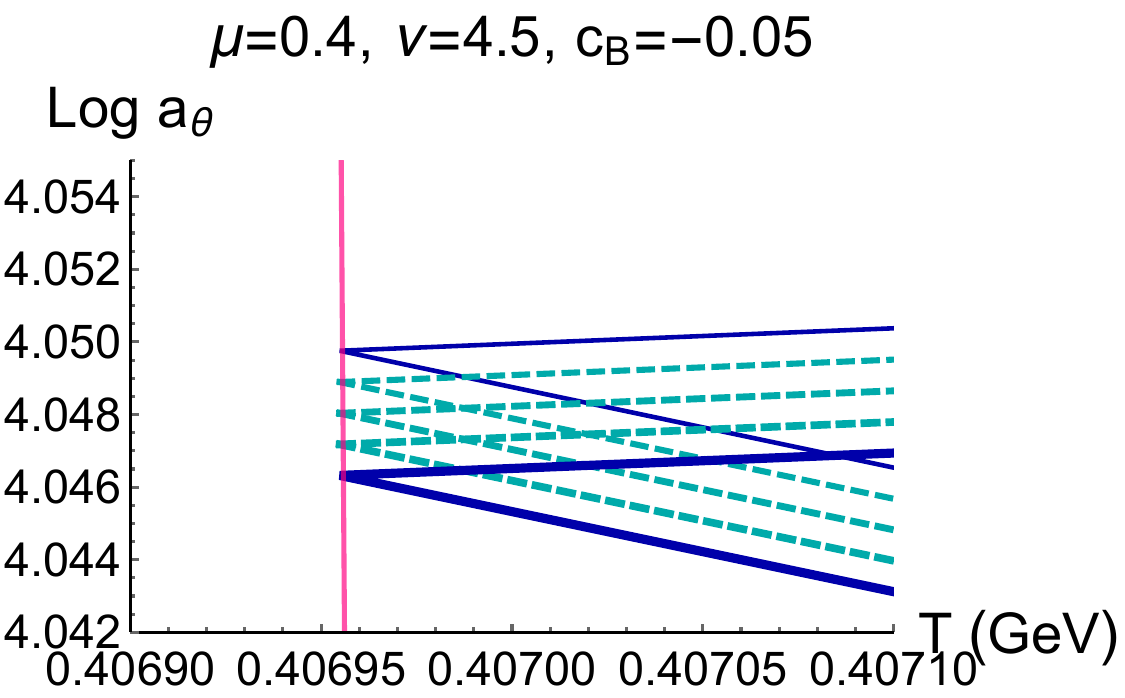} \\ 
A \qquad \qquad \qquad \qquad \qquad \qquad B \qquad \qquad \qquad \qquad \qquad \qquad C
  \caption{ A) The dependence of the JQ parameter on the temperature for different orientation $\theta = 0, \pi/6, \pi/4, \pi/3, \pi/2$ for $\mu=0.4$, $\nu=4.5$ and $c_B=-0.05$ in heavy quarks model. The light magenta line is a jump for the JQ parameter. B) The zoom view of the lower part of (A) near the jump. C) The zoom view of the top part of (A) near the jump. \label{Htheta04}}
\end{figure}

\subsection{Numerical results} \label{3.2}

The numerical results for the JQ parameter are presented. We set $L=1$, take $x_i=x_1$, and study the dependence on the angle $\theta$ in the $(x_2,x_3)$-plane. 

Figure \ref{Htheta1}A and B shows the temperature dependence of the JQ parameter for orientations $\theta = 0, \pi/6, \pi/4, \pi/3, \pi/2$ in the heavy quarks model. Figure \ref{Htheta04} A, B and C display the temperature dependence near the first-order phase transition. A qualitatively similar $\theta$-dependence is observed in the light quarks model.

\section{Conclusion} \label{4}

The JQ parameter is studied in the anisotropic holographic model under a strong magnetic field for light and heavy quarks models.
These holographic models are constructed using the Einstein-dilaton-three-Maxwell action and incorporate a five-dimensional metric, extending earlier isotropic and
partially anisotropic backgrounds. The JQ parameter exhibits discontinuities at the first-order phase transition; the size and location of these jumps depend on the orientation of the lightlike Wilson loop in the anisotropic background.

The important goal for the future consideration is to construct hybrid model that incorporates both heavy and light quarks within a single holographic setup using a unified warp factor. It would be interesting to study the phase transition structure of this model and the JQ parameter behavior near the transitions.

\begin{acknowledgments}
Author would like to
thank the organizers of the XXV International Workshop-School
High Energy Physics and Quantum Field Theory (QFTHEP'270) for the opportunity to present this report. Author would
like to thank Prof. I.Ya. Aref’eva, A. Hajilou, and A. Nikolaev for valuable and useful discussions. 
\end{acknowledgments}

\section*{FUNDING}
The work was supported by Theoretical Physics and Mathematics Advancement Foundation “BASIS (grant No. 23-1-4-43-1).

\section*{CONFLICT OF INTEREST}

The author of this work declares that he has no conflicts of interest.
 


\begin{thebibliography}{}

\bibitem{1}
H.~Liu, K.~Rajagopal and U.~A.~Wiedemann,
Phys. Rev. Lett. \textbf{97}, 182301 (2006)
doi:10.1103/PhysRevLett.97.182301

\bibitem{2}
  J.~Casalderrey-Solana, H.~Liu, D.~Mateos, K.~Rajagopal and
  U.~A.~Wiedemann, Gauge/String Duality, Hot QCD and Heavy Ion Collisions,
 (Cambridge University Press (2014))
  doi:10.1017/CBO9781139136747;

\bibitem{3}
  I.~Ya.~Aref'eva,
  Phys.~Usp. {\bf 57}, 527 (2014), doi:10.3367/UFNe.0184.201406a.0569;

\bibitem{4}
  O.~DeWolfe, S.~S.~Gubser, C.~Rosen and D.~Teaney,
  Prog. Part. Nucl. Phys. {\bf 75}, 86 (2014)

\bibitem{5}
  O.~Andreev and V.~I.~Zakharov,
  Phys.~Rev.~D \textbf{74}, 025023 (2006), doi:10.1103/PhysRevD.74.025023

\bibitem{6}
I.~Y.~Aref'eva and A.~A.~Golubtsova,
JHEP \textbf{04}, 011 (2015), doi:10.1007/JHEP04(2015)011
 
\bibitem{7}
  Y.~Yang and P.~H.~Yuan,
  JHEP \textbf{12}, 161 (2015), doi:10.1007/JHEP12(2015)161

\bibitem{8}
M.~W.~Li, Y.~Yang and P.~H.~Yuan,
Phys. Rev. D \textbf{96}, no.6, 066013 (2017)
doi:10.1103/PhysRevD.96.066013

\bibitem{9}
  I.~Ya.~Aref'eva and K.~A.~Rannu,
  JHEP {\bf 05}, 206 (2018), doi:10.1007/JHEP05(2018)206
  
\bibitem{10}
  H.~Bohra, D.~Dudal, A.~Hajilou and S.~Mahapatra,
  Phys.~Lett.~B \textbf{801}, 135184 (2020), doi:10.1016/j.physletb.2019.135184

\bibitem{11}
 I.~Y.~Aref'eva, K.~Rannu and P.~Slepov,
JHEP \textbf{06}, 090 (2021),
doi:10.1007/JHEP06(2021)090

\bibitem{12}
I.~Y.~Aref'eva, K.~Rannu and P.~Slepov,
JHEP \textbf{07}, 161 (2021)
doi:10.1007/JHEP07(2021)161

\bibitem{13}
I.~Y.~Aref'eva, A.~Ermakov, K.~Rannu and P.~Slepov,
Eur. Phys. J. C \textbf{83}, no.1, 79 (2023)
doi:10.1140/epjc/s10052-022-11166-3

  \bibitem{14}
I.~Y.~Aref'eva, A.~Hajilou, K.~Rannu and P.~Slepov,
Eur. Phys. J. C \textbf{83}, no.12, 1143 (2023)
doi:10.1140/epjc/s10052-023-12309-w

\bibitem{15}
P.~Slepov,
Int. J. Mod. Phys. A \textbf{39}, no.35, 2443015 (2024)
doi:10.1142/S0217751X24430152

\bibitem{16}
I.~Y.~Aref'eva, A.~Hajilou, P.~Slepov and M.~Usova,
Phys. Rev. D \textbf{110}, no.12, 126009 (2024)
doi:10.1103/PhysRevD.110.126009
[arXiv:2402.14512 [hep-th]].

\bibitem{17}
  F.~R.~Brown et al.,
  Phys.~Rev.~Lett. \textbf{65}, 2491-2494 (1990), doi:10.1103/PhysRevLett.65.2491

\bibitem{18}
D.~S.~Ageev, I.~Y.~Aref'eva, A.~A.~Golubtsova and E.~Gourgoulhon,
Nucl. Phys. B \textbf{931}, 506-536 (2018)
doi:10.1016/j.nuclphysb.2018.04.016

\bibitem{19}
S.~Heshmatian and R.~Morad,
``QGP probes from a dynamical holographic model of AdS/QCD,''
Eur. Phys. J. C \textbf{84} (2024) no.4, 360
doi:10.1140/epjc/s10052-024-12596-x

\bibitem{20}
Z.~R.~Zhu and D.~Hou,
Phys. Rev. D \textbf{110}, no.6, 066010 (2024)
doi:10.1103/PhysRevD.110.066010

  \bibitem{21}
  I.~Aref'eva, K.~Rannu and P.~Slepov,
  Phys.~Lett.~B \textbf{792}, 470-475 (2019), doi:10.1016/j.physletb.2019.04.012

\bibitem{22}
  I.~Y.~Aref'eva, A.~Patrushev and P.~Slepov,
  JHEP \textbf{07}, 043 (2020), doi:10.1007/JHEP07(2020)043

\bibitem{23}
I.~Y.~Aref'eva, K.~A.~Rannu and P.~S.~Slepov,
Teor. Mat. Fiz. \textbf{206}, no.3, 400-409 (2021)
doi:10.1134/S0040577921030077

\bibitem{24}
I.~Y.~Aref'eva, K.~Rannu and P.~Slepov,
[arXiv:2012.05758].

\bibitem{25}
P.~S.~Slepov,
Moscow Univ. Phys. Bull. \textbf{79}, no. Suppl 1, 555-558 (2024)
doi:10.3103/S0027134924701480

\bibitem{26}
I.~Y.~Aref'eva, A.~Hajilou, A.~Nikolaev and P.~Slepov,
[arXiv:2507.19426 [hep-th]].

\bibitem{27}
I.~Y.~Aref'eva, K.~Rannu and P.~Slepov,
Theor. Math. Phys. \textbf{222}, no.1, 140-153 (2025)
doi:10.1134/S0040577925010118

\bibitem{28}
P.~S.~Slepov,
Phys. Part. Nucl. \textbf{52}, no.4, 560-563 (2021)
doi:10.1134/S1063779621040572

\end{thebibliography}
\end{document}